# CLUSTERING HIGH DIMENSIONAL DATA USING SUBSPACE AND PROJECTED CLUSTERING ALGORITHMS

Rahmat Widia Sembiring[1], Jasni Mohamad Zain[2], Abdullah Embong[3]

Faculty of Computer Systems and Software Engineering Universiti Malaysia Pahang

Lebuhraya Tun Razak, 26300, Kuantan, Pahang Darul Makmur, Malaysia

email : [1]rahmatws@yahoo.com,  [2]jasni@ump.edu.my,  [3]embonga@gmail.com

*ABSTRACT:*

*Problem statement: Clustering has a number of techniques that have been developed in statistics, pattern recognition, data mining, and other fields. Subspace clustering enumerates clusters of objects in all subspaces of a dataset. It tends to produce many over lapping clusters. Approach: Subspace clustering and projected clustering are research areas for clustering in high dimensional spaces. In this research we experiment three clustering oriented algorithms, PROCLUS, P3C and STATPC. Results: In general, PROCLUS performs better in terms of time of calculation and produced the least number of un-clustered data while STATPC outperforms PROCLUS and P3C in the accuracy of both cluster points and relevant attributes found. Conclusions/Recommendations: In this study, we analyze in detail the properties of different data clustering method.*

## 1. INTRODUCTION

Clustering is concerned with grouping together objects that are similar to each other and dissimilar to the objects belonging to other clusters [1]. Cluster is used to group items that seem to fall naturally together [2]. Various types of clustering: hierarchical (nested) versus partitioned (un-nested), exclusive versus overlapping versus fuzzy, and complete versus partial [3]. Clustering is an unsupervised learning process that partitions data such that similar data items grouped together in sets referred to as clusters. This activity is important for condensing and identifying patterns in data [4].

Clustering technique is applied when there is no class to predict but rather when the instances divide into natural groups. These clusters presumably reflect some mechanism at work in the domain. That causes some instances to bear a stronger resemblance to each other than they do to the remaining instances. Clustering naturally requires different techniques to the classification and association learning methods we have considered so far [2]. Subspace clustering and projected clustering are recent research areas for clustering in high dimensional spaces. However, in high dimensional datasets, traditional clustering algorithms tend to break down both in terms of accuracy, as well as efficiency, so-called curse of dimensionality [5].

This paper will study three algorithms used for clustering. PROCLUS is focused on a method to find clusters in small projected subspaces for data of high dimensionality. It presents an effective method for finding regions of greater density in high dimensional data in a way which has good scalability and usability [6]. P3C is an algorithm for projected clustering that can effectively discover projected clusters in the data while minimizing the number of required parameters. P3C positions itself between projected and subspace clustering in that it can compute both disjoint and overlapping clusters. P3C is the first projected clustering algorithm for both numerical and categorical data [5]. STATPC is an approximation algorithm that aims at extracting from the data axis-parallel regions that "stand out" in a statistical sense. Intuitively, a *statistically significant* region is a region that contains significantly more points than expected [7]. OpenSubspace, is an open source framework that meets these requirements. OpenSubspace integrates state-of-the-art performance measures and visualization techniques to foster research in subspace and projected clustering [8].

This paper will be organized into a few sections. Section 2 will present current work on subspace clustering. Our proposed experiment will be discussed in Section 3. Section 4 will discuss the results by comparing performance of the three algorithms, and followed by concluding remarks in Section 5.

## 2. RELATED WORK

Clustering has been used extensively as a primary tool for data mining, but do not scale well to cluster high dimensional data sets in terms of effectiveness and efficiency, because of the inherent sparsity of high dimensional data. Problem arises when the distance between any two data points becomes almost the same [5], therefore it is difficult to differentiate similar data points from dissimilar ones. Secondly, clusters are embedded in the subspaces of the high dimensional data space, and different clusters may exist in different subspaces of different dimensions [9]. Techniques for clustering high dimensional data have included both feature transformation and feature selection techniques [10].

Density based clustering differentiates regions which have higher density than its neighbourhood and does not need the number of clusters as an input parameter. Regarding a termination condition, two parameters indicate when the expansion of clusters should terminate: given the radius of the volume of data points to look for, $\varepsilon$, a minimum number of points for the density calculations, $\wp$, has to be exceeded [11]. For a broad range of data distribution and distance measure, the relative contrast does diminish as the dimensionality increase [12].

As known, no meaningful cluster analysis is possible unless a meaningful measure of distance or proximity between pairs of data points have been established. Most of the clusters can be identified by their location or density characters [13]. There is a general categorization for high dimensional data set clustering: dimension reduction, parsimonious models, and subspace clustering. A cluster is a dense region of points, which is separated by low-density regions, from other regions of high density. This definition is more often used when the clusters are irregular or intertwined, and when noise and outliers are present [14].

Distance functions have been used in various dimensional clustering algorithms, depending on the particular problem being solved. Manhattan segmental distance is used in PROCLUS that is defined relative to some set of dimension [5]. Employing the segmental distance as opposed to the traditional Manhattan distance is useful when comparing points in two different clusters that have varying number of dimension, because the number of dimension has been normalized. Existing projected clustering algorithms are either based on the computation of $k$ initial clusters in full dimensional space, or leverage the idea that clusters with as many relevant attributes as possible are preferable. Consequently, these algorithms are likely to be less effective in the practically most interesting case of projected clusters with very few relevant attributes, because the members of such clusters are likely to have low similarity in full dimensional space [5].

Several subspace clustering algorithms attempt to compute a succinct representation of the numerous subspace clusters that they produce, by reporting only the highest dimensional subspace clusters, merge similar subspace clusters, or organize them hierarchically. In PROCLUS, algorithms start by choosing a random set of *k* medoid from M and progressively improve the quality of medoid by iteratively replacing the bad medoids in the current set with new point from M [5]. P3C (Projected Clustering via Cluster Cores) effectively discovers the projected clusters in the data while being remarkably robust to the only parameter that it takes as input. Setting this parameter requires little prior knowledge about the data, and, in contrast to all previous approaches, there is no need to provide the number of projected clusters as input, since algorithm can discover, under very general conditions, the true number of projected clusters [12]. In DOC, a mathematical formulation for the notion of optimal projective cluster based on the density of the points in the subspaces is proposed [15]. While SCHISM, which is based on the GenMax algorithm that mines maximal item sets, uses a depth-first search with backtracking to mine the maximal interesting subspaces [16].

## 3. EXPERIMENT RESULT

In this study we run Opensubspace [9] embedded in Weka. We had experiment three clustering oriented methods to optimize the overall clustering result. In PROCLUS, k-medoid algorithm iteratively refining a full-space k-medoid clustering. P3C combines one-dimensional cluster cores to higher-dimensional cluster. STATPC uses a statistical test to remove redundant clusters out of the result. OpenSubspace integrates state-of-the-art performance measures and visualization techniques to foster research in subspace and projected clustering. We use a synthetic data course implemented. After setting a required parameter for PROCLUS we obtain the results as follows. Setup parameter was done at subspace cluster bracketing, and average dimension and number of cluster were defined. Visualization of the number of clusters is shown in Figure 1(b). Figure 1(c), shows plot of matrix for all of attribute with data record. Figure 1(d), shows visualization one of attribute SCU01 related to SP attribute.

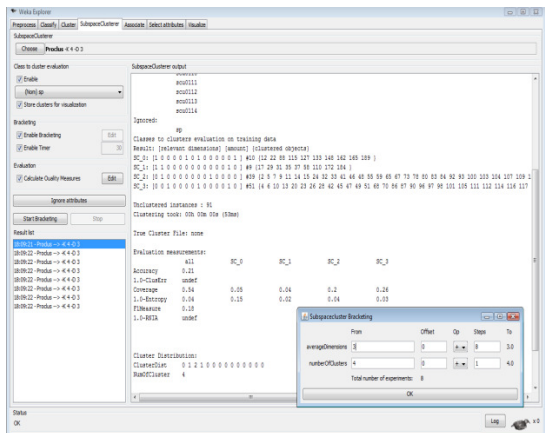

*Figure 1(a): Subspace Cluster Output for PROCLUS*

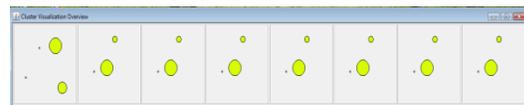

*Figure 1(b): Number of Cluster for PROCLUS output*

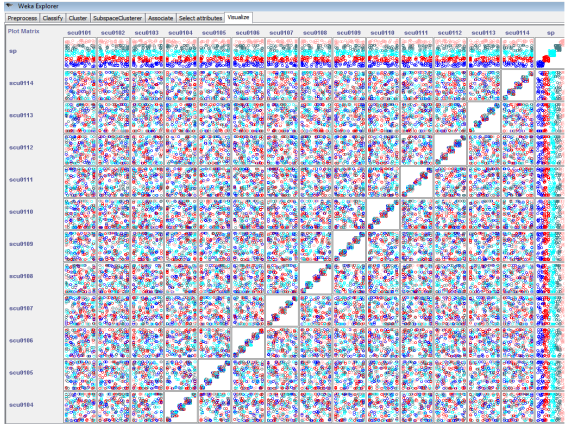

*Figure 1(c): Plot of matrix for PROCLUS clustering method*

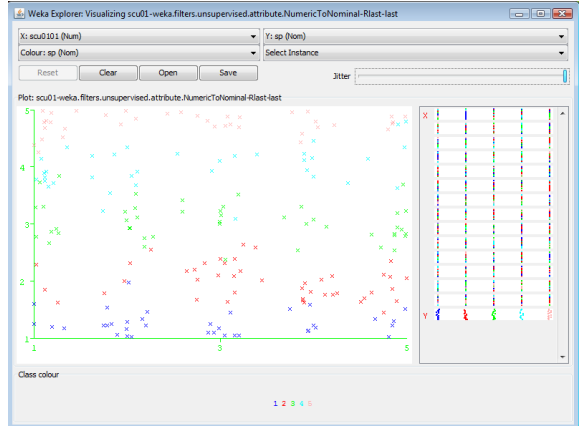

*Figure 1(d): Visualization one of attribute for PROCLUS clustering method*

After setting a required parameter for P3C we have obtained results as follows. Setup parameter was done at subspace cluster bracketing, and average dimension and number of cluster were defined. Visualization of number of clusters is shown in Figure 2(b). Figure 2(c), shows plot of matrix for all attributes with data record. Figure 2(d), shows the visualization one of attribute SCU01 related to SP attribute.

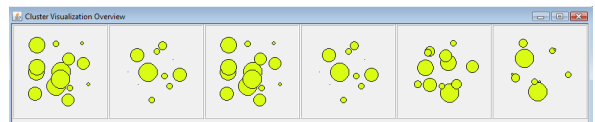

*Figure 2(b): Number of Clusters for P3C output*

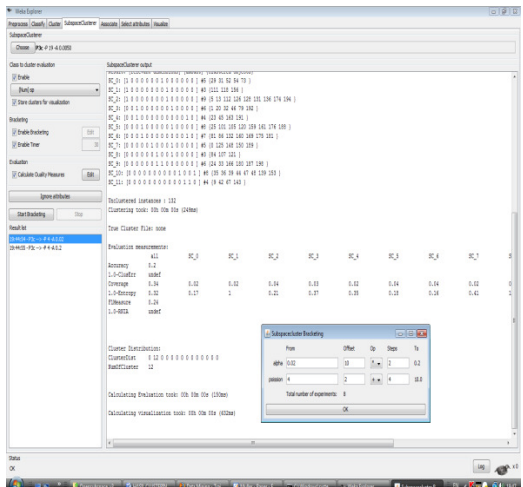

*Figure 2(a): Subspace Cluster Output for P3C*

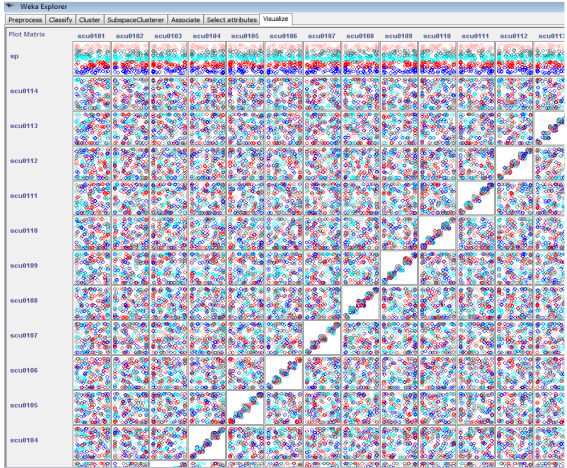

Figure 2(c): Plot of matrix for P3C clustering method

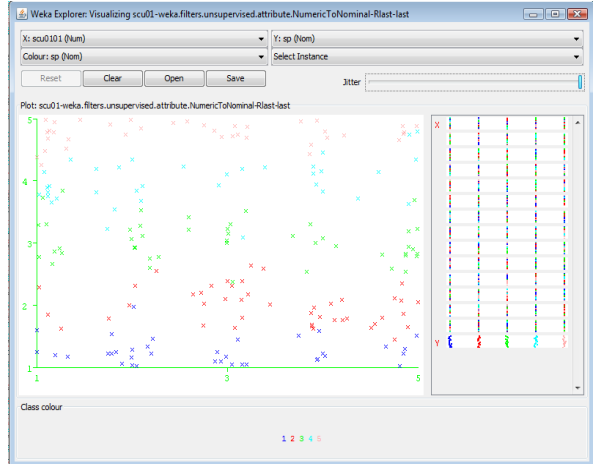

Figure 2(d): Visualization one of attribute for PROCLUS clustering method

After setting a required parameter for STATPC we have obtained results as follows. Setup parameter was done at subspace cluster bracketing, and average dimension and number of cluster were defined. Visualization of number of clusters is shown in Figure 3(b). Figure 3(c), shows plot of matrix for all of attribute with data record. Figure 3(d), shows the visualization one of attribute SCU01 related to SP attribute.

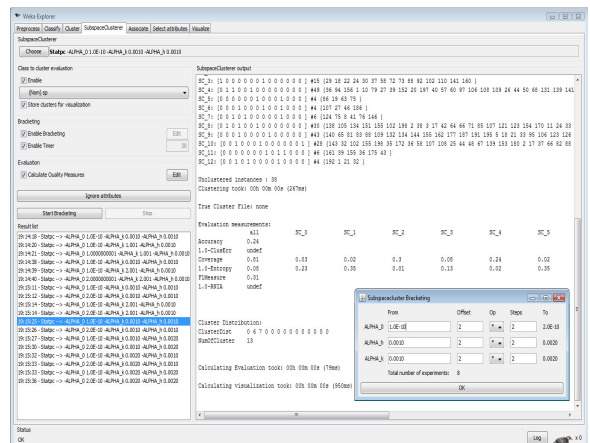

Figure 3(a): Subspace Cluster Output for STATPC

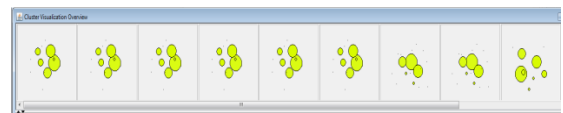

Figure 3(b): Number of Clusters for STATPC output

166

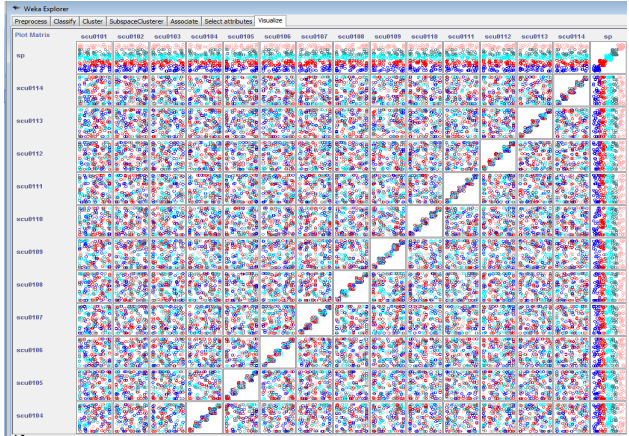
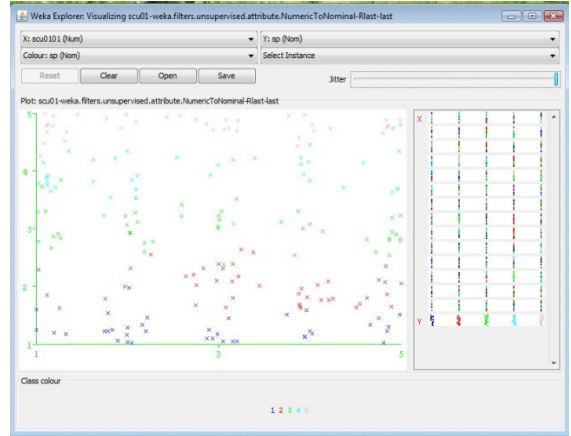

*Figure 3(c): Plot of matrix for P3C clustering method*

*Figure 3(d): Visualization one of attribute for PROCLUS clustering method*

## 4. DISCUSSION

### 4.1 Accuracy Analysis

To ensure good accuracy of the output, PROCLUS was able to achieve two essential results: find a piercing set of medoids, and associate the correct set of dimension to each medoids [5]. On synthetic data, the number of clusters discovered by P3C equals the true number of projected clusters in the data. On numerical data, P3C effectively discovers projected clusters with varying orientation in their relevant subspaces. The accuracy of P3C on datasets where projected clusters have axis-parallel orientation is as high as the accuracy of P3C on datasets where projected clusters have arbitrary orientation [5].

STATPC outperforms previously proposed projected and subspace clustering algorithms in the accuracy of both cluster points and relevant attributes found [5]. Using our synthetic data (university course implemented while industrial training), we found F1, accuracy, entropy and coverage of data between PROCLUS, P3C and STATPC as shown in Figure 4(a). From the figures below, we can see that the accuracy of these algorithms were equal, while PROCLUS has significant F1 value. Meanwhile, for faculty course implemented, we found accuracy of PROCLUS is lower than the other (Figure 4(b)). For faculty course implemented data, we found accuracy of P3C is lower than the other (Figure 4(c)), and value F1 for STATPC is greater than the others.

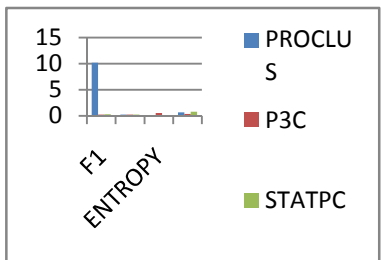
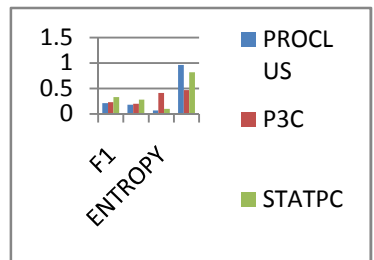
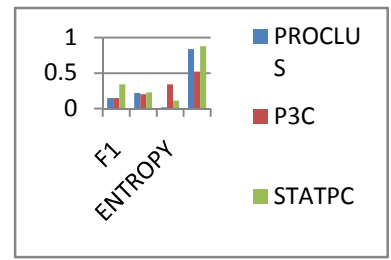

*Figure 4(a): Comparative of measure for university course implemented*

*Figure 4(b): Comparative of measure for faculty course implemented*

*Figure 4(c): Comparative of measure for study program course implemented*

### 4.2 Number of clusters analysis

P3C requires only one parameter setting, namely the Poisson threshold. P3C does not require the user to set the target number of clusters; instead, it discovers a certain number of clusters by itself [5]. On synthetic data, STATPC set the target number of clusters to the number of implanted clusters [5].

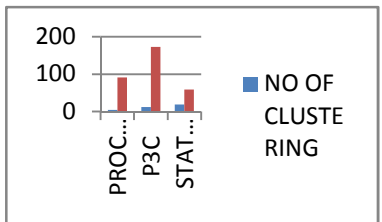

*Figure 5(a): Comparative of clustering result for university course implemented*

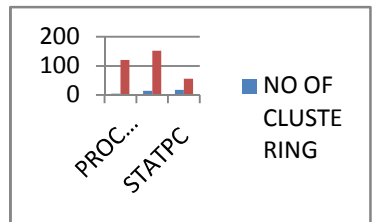

*Figure 5(b): Comparative of number of clusters for faculty course implemented*

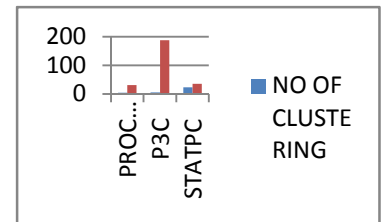

*Figure 5(c): Comparative of number of clusters for study program course implemented*

Using our synthetic data (university course implemented while industrial training), we found the number of clustered and un-clustered data for PROCLUS, P3C and STATPC as shown in Figure 5(a). From figure above, we can see that the number of clusters using PROCLUS was less than the others, but STATPC result has a lower un-clustered data. For faculty course implemented data, we found the number of clusters of PROCLUS is lower than the others (Figure 5(b)), and un-clustered data using P3C is more than the others. For study program course implemented data, we found the number of clusters using PROCLUS is lower than the other (Figure 5(c)), and for un-clustered data, P3C is greater than the others.

### 4.3 Time of Calculation Analysis

STATPC has a longer runtime than previous algorithms. The number of times X that an attribute occurs in a subset of M randomly selected pairs of attributes is a hyper geometric distributed variable [5]. The running time of P3C increases with increasing average cluster dimensionality, due to the increased complexity of signatures generation. However, as the average cluster dimensionality increases, clusters become increasingly detectable in full dimensional space [5]. Using our synthetic data (university course implemented while industrial training), we recorded the time of calculation and the time of visualization between PROCLUS, P3C and STATPC as shown in Figure 6(a).

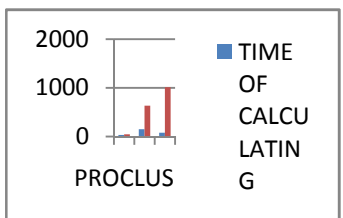

*Figure 6(a): Comparative time of calculation for university course implemented*

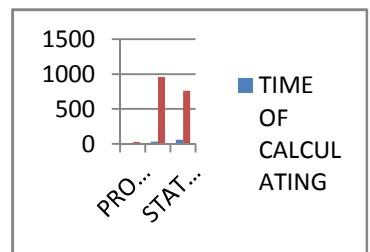

*Figure 6(b): Comparative time of calculation for faculty course implemented*

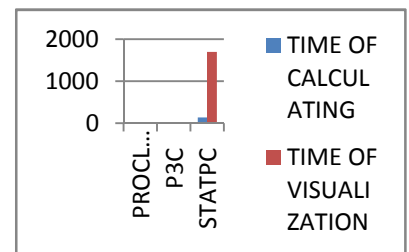

*Figure 6(c): Comparative time of calculation for study program course implemented*

From figure above, we can see the time of calculation and the time of visualization using PROCLUS is faster than the others. For faculty course implemented data, we found that the time of calculation and the time of visualization using PROCLUS were faster than the others (Figure 6(b)). For study program course implemented data, we found the time of calculation and the time of visualization using P3C is faster than others (Figure 6(c)).

## 5. CONCLUSION AND FUTURE WORK

In this work, we discussed existing projected and subspace clustering literature. We have compared between three available algorithms, we found advantages and disadvantages. In general, PROCLUS is better in term of time for calculation and obtained the least number of un-clustered data. STATPC outperforms PROCLUS and P3C in the accuracy of both cluster points and relevant attributes found. In the future we will study cell-based subspace clustering and density-based subspace clustering.

[1] Ali Alijamaat, Madjid Khalilian, Norwati Mustapha, "A Novel Approach for High Dimensional Data Clustering," wkdd, pp.264-267, 2010 Third International Conference on Knowledge Discovery and Data Mining, 2010, [doi>10.1109/WKDD.2010.120]

[2] Witten, Ian H, Eibe Frank, 2005, Data Mining–Practical Machine Learning Tools and Technique, 2nd edn, Morhan Kaufmann, San Fransisco

[3] Tan, Pang Nin, Michael Steinbach,Vipin Kumar, 2006, Introduction to Data Mining, Pearson International Edition, Boston

[4] Poncelet, Pascal, Maguelonne Teisseire, 2008, Florent Masseglia, Data Mining Patterns : New Method and Application, London

[5] Gabriela Moise,, Jorg Sander, Finding Non-Redundant, Statistically Significant Regions in High Dimensional Data : A Novel Approach to Projected and Subspace Clustering, 2008, Proceedings of the 14th ACM SIGKDD International Conference on Knowledge Discovery and Data Mining, Las Vegas, Nevada, USA, August 24-27, 2008 [doi>10.1145/1401890.1401956]

[6] Charu C. Aggarwal, Joel L. Wolf , Philip S. Yu, Cecilia Procopiuc, Jong Soo Park, Fast algorithms for projected clustering, Proceedings of the 1999 ACM SIGMOD international conference on Management of data, p.61-72, May 31-June 03, 1999, Philadelphia, Pennsylvania, United States  [doi>10.1145/304182.304188]

[7] Gabriela Moise, Jorg Sander, Martin Ester, "P3C: A Robust Projected Clustering Algorithm," icdm, pp.414-425, Sixth IEEE International Conference on Data Mining (ICDM'06), 2006, [doi>10.1109/ICDM.2006.123]

[8] Müller E., Günnemann S., Assent I., Seidl T.: Evaluating Clustering in Subspace Projections of High Dimensional Data In Proc. 35th International Conference on Very Large Data Bases (**VLDB 2009**), Lyon, France, http://dme.rwth-aachen.de/OpenSubspace/evaluation/

[9] Gan, Guojun, Jianhong Wu, Subspace Clustering For High Dimensional Categorical Data, 2004, SIGKDD Explorations 2004, Volume 6, p.87-94 [doi>10.1145/1046456.1046468]

[10] Lance Parsons , Ehtesham Haque , Huan Liu, Subspace clustering for high dimensional data: a review, ACM SIGKDD Explorations Newsletter, v.6 n.1, p.90-105, June 2004, [doi>10.1145/1007730.1007731]

[11] Bicici, Ergun, Deniz Yuret, 2007, Local Scaled Density Based Clustering, ICANNGA (1) 2007: 739-748, [doi>10.1.1.91.6240]

[12] Houle, Michael E.; Kriegel, Hans-Peter; Kröger, Peer; Schubert, Erich; Zimek, Arthur (2010), "Can Shared-Neighbor Distances Defeat the Curse of Dimensionality?", Proceedings of the 21th International Conference on Scientific and Statistical Database Management (SSDBM) (Heidelberg, Germany: Springer), [doi>10.1007/978-3-642-13818-8_34]

[13] Wang, Chuang, Ya-yu Lo, Yaoying Xu, Yan Wang, 2007, Constructing the search for a job in academia from the perspective of self-regulated learning strategies and social

cognitive career theory, Journal of Vocational Behavior, v70 n3 p574-589 Jun 2007, [doi>10.1016/j.jvb.2007.02.002]

[14] Steinbach, Michael, Levent Ertöz, and Vipin Kumar, The Challenges of Clustering High Dimensional Data, 2003, In New Vistas in Statistical Physics: Applications in Econophysics, Bioinformatics, and Pattern Recognition, [doi>10.1.1.99.7799]

[15] Procopiuc, C. M., Jones, M., Agarwal, P. K., & Murali, T. M. (2002). A Monte Carlo algorithm for fast projective clustering. Proceedings of the 2002 ACM SIGMOD international conference on Management of data SIGMOD 02. ACM Press, [doi>10.1145/564691.564739]

[16] Karlton Sequeira, Mohammed Zaki, SCHISM : A New Approach to Interesting Subspace, International Journal of Business Intelligence and Data Mining Issue: Volume 1, Number 2 / 2005, p.137-160, [doi>10.1.1.61.1278]